%% file: main.tex
\documentclass{article}
\usepackage{spconf,amsmath,graphicx}

\usepackage{xcolor}

\title{DiffMoog: a Differentiable Modular Synthesizer for Sound Matching}
\namemultiple{Noy Uzrad$^{\star}$ \qquad Oren Barkan$^{\dagger}$ \qquad Almog Elharar$^{\star}$}{Shlomi 
Shvartzman$^{\star}$ \qquad Moshe Laufer$^{\star}$ \qquad Lior Wolf$^{\star}$ \qquad Noam Koenigstein$^{\star}$}

\address{$^{\star}$ Tel-Aviv University, $^{\dagger}$ The Open University, Israel}

\usepackage{url}
\usepackage{amsmath,cite,url}
\usepackage{graphicx}
\usepackage{subfigure}
\usepackage{color}
\usepackage{url}
\usepackage[export]{adjustbox}
\usepackage{hhline}
\usepackage{algorithm}
\usepackage{algorithmic}
\usepackage{setspace}
\newcommand{\modelname}{DiffMoog}

\begin{document}
\ninept
\maketitle

\input{01_Abstract}


\input{02_Introduction}

\input{03_Related_Work}

\input{05_DiffMoog}

\input{07_Experiments}

\clearpage
\newpage
\bibliographystyle{IEEEbib}
\bibliography{strings,refs}

\end{document}

%% file: 01_Abstract.tex
\begin{abstract}
This paper presents \modelname{} - a differentiable modular synthesizer with a comprehensive set of modules typically found in commercial instruments. Being differentiable, it allows integration into neural networks, enabling automated \emph{sound matching}, to replicate a given audio input. Notably, \modelname{} facilitates modulation capabilities (FM/AM), low-frequency oscillators (LFOs), filters, envelope shapers, and the ability for users to create custom signal chains. We introduce an open-source platform that comprises \modelname{} and an end-to-end sound matching framework. This framework utilizes a novel \emph{signal-chain loss} and an encoder network that self-programs its outputs to predict DiffMoogs parameters based on the user-defined modular architecture. Moreover, we provide insights and lessons learned towards sound matching using differentiable synthesis. Combining robust sound capabilities with a holistic platform, \modelname{} stands as a premier asset for expediting research in audio synthesis and machine learning. Our code is released at: \url{https://github.com/aisynth/diffmoog}.
\end{abstract}

%% file: 02_Introduction.tex
\section{Introduction}\label{sec:introduction}
Synthesizers are electronic musical instruments capable of generating a vast spectrum of sounds, from simple tonalities to complex auditory textures, central to many music genres. Composed of modules like oscillators, filters, low frequency oscillators, ADSR envelopes and modulators, they are controlled by an interface to fine-tune the sound output~\cite{russ2004sound}. However, the multitude of interacting parameters makes sound design complex, often demanding deep expertise in sound synthesis and iterative testing.

Neural networks are increasingly used in sound design to replicate input sounds. While initial methods optimized a loss over synthesizer parameters~\cite{barkan2019inversynth, Esling2019, yee-king2018automatic}, it may be beneficial to optimize over the sound directly, since the ultimate goal in sound matching is a high fidelity reproduction. Furthermore, replicating unlabeled out-of-domain sounds not generated by the synth at hand requires unsupervised learning. The implementation of traditional synthesizers does not support automatic differentiation, which is vital for direct sound comparison and optimization via backpropagation of gradients. To address this, several works proposed differentiable synthesizers~\cite{engel2020ddsp, masuda2021synthesizer,caspe2022ddx7}. However they either presented complex~\cite{engel2020ddsp} or overly simplistic~\cite{masuda2021synthesizer} models that deviate from real world synthesizers. For example, both synthesizers from~\cite{engel2020ddsp,masuda2021synthesizer} lack modularity and conventional sound modules (e.g., LFOs, FM/AM modulators), leaving a gap for practical applications.

We present \modelname{} - a differentiable synthesizer with a modular architecture. It incorporates modules commonly found in commercial instruments, following an explainable design familiar to those versed in sound synthesis. The modularity enables both the creation of custom signal chains and the ability to isolate modules for research purposes. \modelname{} pairs with an end-to-end platform, facilitating its integration into an automatic sound matching system, along a newly crafted `signal-chain loss' aimed at guiding the optimization process. The platform supports dataset creation, experiment configuration, model training, and offers comprehensive logging and analysis tools. By releasing \modelname{} as open-source, we aim to propel AI-guided sound synthesis forward. We discuss its implementation, evaluate its potential for sound matching, and share key insights and lessons gained throughout the experimentation process. We note that in the context of sound matching, the sub-task of frequency estimation through gradient descent techniques via minimizing spectrogram-based losses is an intrinsic challenge that remains open~\cite{turian2020sorry}, as we discovered through our own experimentation. Previous studies~\cite{masuda2021synthesizer, caspe2022ddx7, engel2020ddsp} have relied on alleviating assumptions about the data or employed pitch estimation algorithms like CREPE~\cite{kim2018crepe}. However, we did not employ such techniques in our optimization process. 

Our main contributions include: (1) The open-sourced DiffMoog synthesizer and sound matching platform, aiming to provide an easy gateway to conduct research in the field of AI sound synthesis and sound matching. (2) The introduction of a novel signal-chain loss. (3) Lessons and insights learned from optimizing DiffMoog; and (4) Demonstrating the superiority of the Wasserstein loss in frequency estimations.

%% file: 03_Related_Work.tex
\section{Related Work}

\textbf{Non-Differentiable Synthesizer Sound Matching:}
With the rise of machine learning methods, works in sound matching, also known as \emph{parameters inference}\cite{Esling2019}, utilized supervised datasets of sound samples and their parameters, derived from typical non-differentiable synthesizers. Using this data, neural networks were trained to predict sound parameters, a method seen in commercial VST instruments~\cite{yee-king2018automatic,Itoyama2014Parameter,Esling2019}. Sound matching was also explored on custom synthesizers~\cite{barkan2019deep,barkan2019inversynth}. While supervised methods optimize over synthesizer parameters, the ultimate goal of sound matching  is obtaining a high-fidelity sound reconstruction, suggesting a direct audio optimization. These methods are also bound to synthesizer-specific data.  \modelname{} bypasses these restrictions, being differentiable, allowing direct input-output sound comparisons beyond parameter reliance, broadening training and evaluation scope to include out-of-domain sounds. Noteworthy, direct sound optimization can also be achieved using genetic algorithms (GA)~\cite{yee-king2018automatic, tatar2016automatic}. However, GAs are inefficient, necessitating a vast number of synthesizer renders for a single match.

\textbf{Neural Network Based Synthesizers:} Advancements in deep learning gave rise to a suite of neural network based synthesizers~\cite{Engel2017,defossez2018sing,engel2019gansynth,kong2020diffwave, caillon2021rave, le2021improving}. Unlike traditional methods relying on the superposition of sinusoids or the routing of audio submodules (e.g., Additive, Subtractive, FM, Wavetable synthesis and their variants), these synthesizers generate sound directly as the output of a neural network. However, they lack the control and explainability that the classic synthesizers offer. Other works explored the power of generative adverserial networks . GANSynth~\cite{engel2019gansynth, chandna2019wgansing} and diffusion models~\cite{kong2020diffwave, huang2023noise2music}.
Setting itself apart, DiffMoog uses traditional synthesizer modules, granting full user control and explainable behavior.

\textbf{Differentiable DSP:} 
Differentiable digital signal processing (DDSP)~\cite{engel2020ddsp} integrates signal processing modules as differential operations into neural networks, allowing backpropagation. It uses a flexible additive synthesizer that can produce complex, realistic sounds. Additive synthesis is based on the Fourier theorem of stacking sinusoidal waves to construct a complex sound. A DDSP-inspired wavetable synthesizer was introduced in~\cite{shan2022differentiable}, where arbitrary waveshapes are stacked to generate the final sound. Both methods have limited control due to the large number of parameters with unpredictable effect on the final sound, though DDSP does allow manipulation of fundamental frequency and sound volume.
Differentiable methods had also been employed in audio effects applications~\cite{kuznetsov2020differentiable,ramirez2021differentiable}. Recent advancements include a differentiable mixing console for automatic multitrack mixing~\cite{steinmetz2021automatic}, transferring audio effects and production style between recordings through differentiable effects~\cite{steinmetz2022style}, and automating DJ transitions with differentiable audio effects~\cite{chen2022automatic}.

\textbf{Differentiable Synthesizers:}
Differentiable synthesizers aim to model commercial synthesizer characteristics, often using subtractive synthesis, in which an initial harmonically rich oscillator is further manipulated to modify its frequency content. A basic synthesizer, built upon two additive oscillators and a filter is used in~\cite{masuda2021synthesizer}. Recently, the same authors enhanced their synthesizer to include ADSR envelopes and effects~\cite{masuda2023synthesizer}. Yet, they lack certain complexities found in standard instruments, like LFOs, FM modules, and flexible routing options. FM synthesis as advised by Chowning~\cite{chowning1973synthesis}, uses sine wave oscillators that are interconnected such that they serve as carriers and modulators, controlled by ADSR envelopes that can create complex harmonics and timbres evolving in time. Caspe et al.~\cite{caspe2022ddx7}, manually implemented a few differential patch configurations from the well-known DX7 FM synthesizer~\cite{lavengood2019makes}. Our proposed \modelname{}, to the best of our knowledge, is the first and most comprehensive modular differentiable synthesizer of its kind. It integrates both FM and subtractive synthesis techniques, offering unparalleled capabilities for sound generation and customization within the domain of differentiable synthesis.

%% file: 05_DiffMoog.tex
\section{The DIFFMOOG Synthesizer and Sound Matching Platform}

For clarity, we have aligned the nomenclature in this paper with the terms in our codebase. The code is organized as follows: \\(1) Core implementation of the synthesizer - \texttt{src/synth}. \\(2) Functionality of the sound matching system - \texttt{src/model}.\\ (3) Operations related to dataset management - \texttt{src/dataset}.\\(4) Configurations for training, synthesis, and loss functions - \texttt{configs} directory.

\subsection{\modelname{}}
\modelname{} is a modular synthesizer, inspired by Robert Moog's seminal Moog Synthesizer of 1964, and stands as a differentiable synthesizer, optimized for gradient-based computations. \modelname{} integrates the qualities of subtractive modular synthesis with rich-harmonic oscillators, while also offering the capabilities of FM and additive synthesis. Through its versatile array of modules—oscillators, LFOs, filters, modulators, and shapers — \modelname{} is capable of producing a wide range of soundscapes surpassing previous differentiable synthesizers.

\textbf{The architecture} is built upon a 2D \emph{matrix} of \emph{cells}. These cells act as containers for the synthesizer \emph{modules}  and their associated \emph{parameters}, as defined in the \texttt{src/synth/synth\_architectur-\\e.py} file. This matrix is structured by channels (rows) and layers (columns). Each cell, uniquely identified by its channel and layer indices, can host a single module. 
Signal flow within this structure is managed by: (1) \emph{connections} which are assets of the cells: each cell is allowed to have multiple input and output connections, which dictate how signals traverse the matrix, and (2) using activation parameters to control module behavior, ranging from a simple on/off switch to bypassing an internal procedure. For example, one could opt to use just one oscillator where two are present or neutralize FM in a modulator.
In order to simplify the signal flow and precisely define the sound generation process described later, the architecture imposes a rule on cell connections: a cell can only receive input from cells in preceding layers and send outputs to subsequent layers. 

\begin{figure}
\setlength{\fboxsep}{0pt}
 \centerline{
 \includegraphics[width=0.8\linewidth, height=0.45\linewidth]{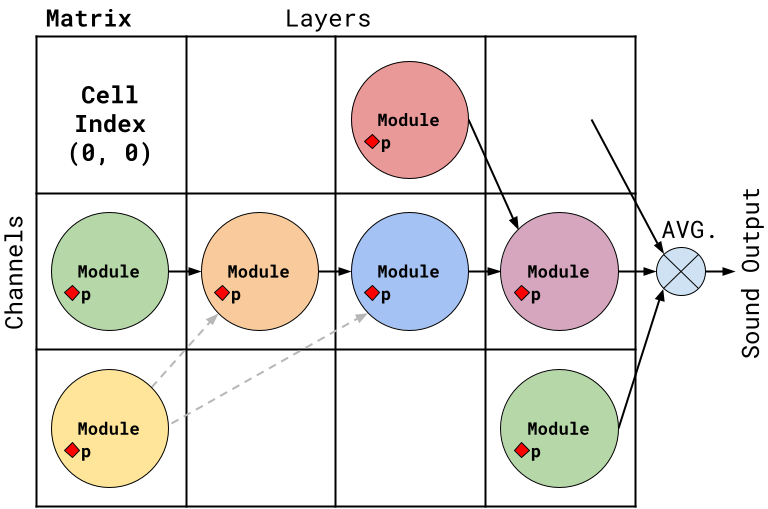}}\vspace{-2mm}
 \caption{The \modelname{} synth with an arbitrary \emph{chain}. Shown: \emph{matrix}, \emph{cells}, \emph{modules}, \emph{connections} (arrows) and \emph{parameters} (`p'). Black arrows are \emph{fixed connections}, gray arrows are \emph{optional connections}. Empty cell outputs the 0 signal.}\vspace{-4mm}
 \label{fig:diffmoog}
\end{figure}

\modelname{} allows users to craft signal chains using predefined modules and connections, collectively termed \emph{chains}, detailed in \texttt{src/synth/synth\_chains.py}. This feature enables versatile synthesizer configurations, but also permits module isolation for research needs. Utilizing a particular chain, one can generate random datasets. Within a chain, connections can be (1) \emph{fixed}, always present, or (2) \emph{optional}, which, during dataset creation, are decided upon randomly for each sound instance in conjunction with the module's activation parameters to finalize the signal flow. Noteworthy, sound modules may possess optional inputs, and these must correspond with connections leading to their respective cells. An arbitrary chain of the \modelname{} synthesizer is depicted in Figure~\ref{fig:diffmoog}.

After populating the synthesizer matrix with modules and defining the cells interconnections (i.e. setting up a chain), the signal chain and system state are solidified. Sound generation proceeds layer-by-layer, from the lowest to the highest order. In each layer, cells compute sound independently, processing their (optional) input signals via their module logic and parameters. The signals from the cells then serve as input for cells in subsequent layers. Upon completion, the final layer's outputs across all channels are averaged to yield the final sound output (See \texttt{src/synth/synth\_architecture.py::SynthModula-\\r.generate\_signal()}).

\textbf{The synthesizer modules}, are implemented in \texttt{src/synth/s-\\ynth\_modules.py}. The Oscillator generates sine, square, and sawtooth signals spanning the human hearing range (20Hz-20KHz), employing closed-form expressions instead of the additive synthesis used in previous works like \cite{masuda2021synthesizer}. The Low-Frequency Oscillator (LFO) acts as a sub-audible control signal, enriching temporal variations in sound, especially in the FM and Tremolo contexts. Furthermore, the system integrates FM Oscillator for frequency modulation, and a Filter typical for subtractive synthesis. Control of amplitude and filter characteristics over time is facilitated by the ADSR modules, while the Mix module blends input sounds, and Tremolo modulates amplitude for a pulsing effects.

\begin{figure}[t]
  \centering
  \subfigure[AM Square wave using tremolo effect ]{\includegraphics[width=0.49\linewidth, height=2.5cm]{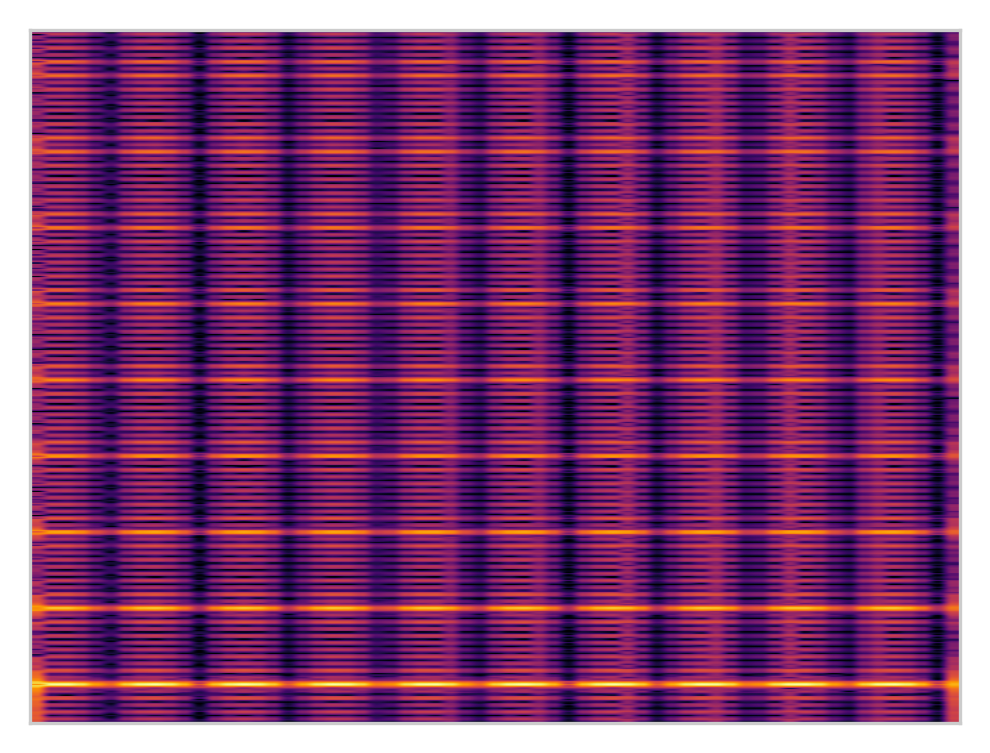}}\hfill
  \subfigure[FM square modulated by a sawtooth wave]{\includegraphics[width=0.49\linewidth, height=2.5cm]{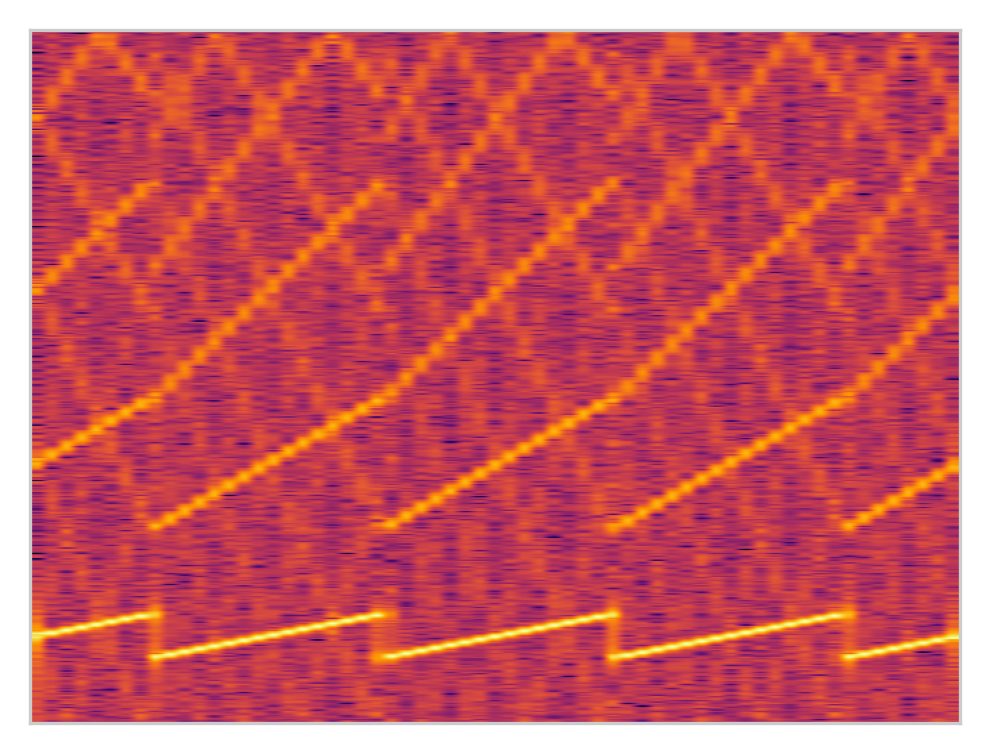}}\hfill\vspace{-3mm}
  \caption{Spectrograms sounds synthesized with \modelname{}. While typical, FM/AM sounds cannot be synthesized by prior differentiable synths.}\vspace{-2mm}
  \label{fig:plots}
\end{figure}

Spectrogram plots of several sound examples generated by the \modelname{} synthesizer are shown in Figure~\ref{fig:plots}. The corresponding audio examples are available for listening on the DiffMoog's GitHub repository. 

\subsection{End-To-End Sound Matching Platform}

\begin{figure}
\setlength{\fboxsep}{0pt}
 \centerline{
 \includegraphics[width=0.85\linewidth, height=0.35\linewidth]{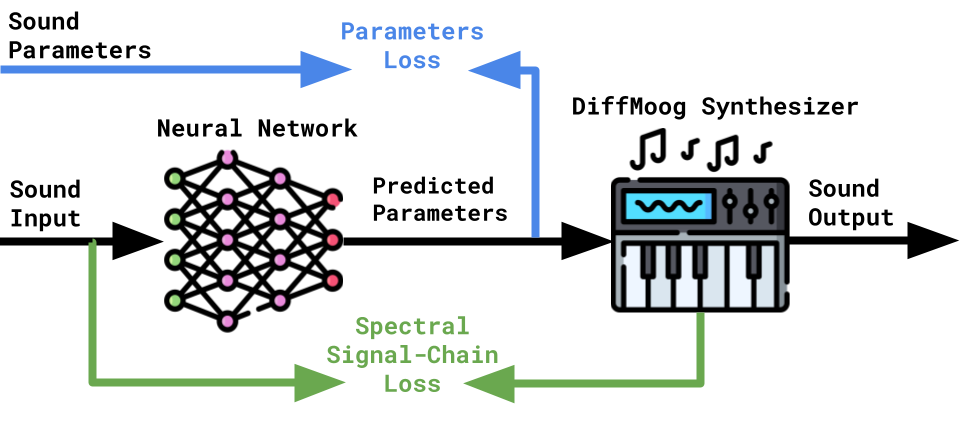}}\vspace{-3mm}
 \caption{The end-to-end sound matching system diagram.}\vspace{0mm}
 \label{fig:system}
\end{figure}

To train a model, an audio input is fed into a neural network, which predicts parameters for modules in a specified \modelname{} chain. In turn, the predicted parameters are used to synthesize a sound output that is optimized to replicate the input. This process is guided by minimizing both spectral and parameter losses. Since the parameter loss is optional, the system allows unsupervised training of unlabeled sounds, including out-of-domain sounds not made by \modelname{}. At inference time, given a new audio input, the neural network outputs parameter predictions for the corresponding \modelname{} chain that can be used to synthesize the predicted sound. A system diagram is shown in Fig.~\ref{fig:system}.

\begin{figure}
\setlength{\fboxsep}{0pt}
 \centerline{
 \includegraphics[width=1\linewidth, height=2cm]{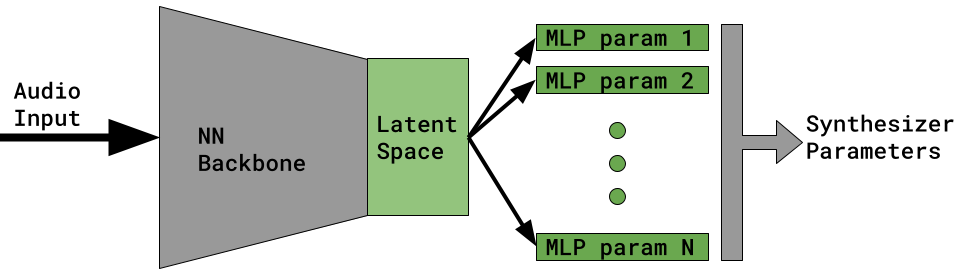}}
 \caption{The neural network architecture with dynamically allocated MLP heads}\vspace{-1mm}
 \label{fig:nn}
\end{figure}

\textbf{The encoder network} features a neural backbone (e.g., ResNet, GRU) that produces a latent vector, complemented by a dynamically configured Multi-Layer Perceptron (MLP) head for each parameter, tailored to the selected synth chain. The encoder architecture is illustrated in Fig.~\ref{fig:nn}. Notably, connections are not predicted as they are assets of the cells. Instead, the signal flow is governed exclusively by classified activation parameters. During training, connections deemed as 'optional' within the chain are treated as 'fixed'. Continuous parameters are normalized using the sigmoid function, while the `Gumbel-Softmax'~\cite{jang2016categorical} trick is employed for sampling from categorical parameters. The reader is referred to \texttt{src/model/model.py} in our repository for the exact code implementation

\textbf{The loss function} we employed is a combined loss consisting of two terms: a parameters loss and a newly proposed spectral loss we name \emph{signal-chain loss}. The \textbf{parameters Loss}, denoted as $\mathcal{L}_{\mathrm{p}}$, is the sum of synthesizer parameters differences between predicted and original values. Regression parameters employ L1 or L2 loss, while categorical parameters use cross-entropy loss. The overall parameters loss is formulated as follows:
\begin{equation}
\small
\mathcal{L}_{\mathrm{p}} = \sum_{\mathrm{n} \in \mathcal{N}} L_{\text{reg}}(p_n, \hat{p}_n) + \sum_{\mathrm{m} \in \mathcal{M}} L_{\text{cat}}(c_m, \hat{c}_m)
\label{eq:loss_parameters}
\end{equation}
where regression and categorical parameters belong to the sets  $\mathcal{N}$ and $\mathcal{M}$ respectively, $r_n$ and $c_m$ represent the genuine values, while $\hat{r}_n$ and $\hat{c}_m$ stand for the predicted values. The loss functions for these parameter types are symbolized by $L{_\text{reg}}$ and $L{_\text{cat}}$, respectively.
The \textbf{signal-chain loss} measures the difference between the ground truth and predicted audio signals. However, unlike prior works, it evaluates the signal at \textbf{all stages} within the synthesizer signal-chain, not only at the final output. Given that the synthesizer's output can be seen as a function composition (e.g., $y = f(g(x))$ where $y$ is the output signal, $f$ and $g$ are sound modules and $x$ is an input signal), it aims to guide the optimization by improving early stage predictions. The signal-chain loss $\mathcal{L}_{\mathrm{SC}}$ is defined using Eq. \ref{eq:loss_sc} and \ref{eq:loss_ifs} below :
\begin{equation}
\small
\mathcal{L}_{\mathrm{SC}} = \sum_{\mathrm{i} \in \mathcal{I}} \sum_{\mathrm{j} \in \mathcal{J}} \sum_{\mathrm{k} \in \mathcal{K}} L_{ijk}
\label{eq:loss_sc}
\end{equation}
\begin{equation}
L_{ijk} = \left\lVert (F_k(S_{ij}(x)) - F_k(S_{ij}(\hat{x}))) \right\rVert_{p}
\label{eq:loss_ifs}
\end{equation}
where indices $\mathrm{i}$, $\mathrm{j}$, and $\mathrm{k}$ indicating cell, FFT window size, and processing type, respectively. The processing function $F_\mathrm{k}$ can be identity, log, or cumulative sum along time/frequency axes (Wasserstein). The $p$ norm is either 1 or 2, and $S$ denotes a spectrogram or mel-spectrogram transform. 
Importantly, the sets $\mathcal{I}, \mathcal{J}, \mathcal{K}$ are configurable, allowing various loss configurations e.g. multiresolution comparison, using traditional output-only optimization, etc. In total, the combined loss is defined as:
\begin{equation}
\small
L_{\mathrm{total}} = \mathcal{L}_{\mathrm{p}} + \beta \cdot
\mathcal{L}_{\mathrm{SC}}
\label{eq:loss_total}
\end{equation}
with $\beta$ as a weighting factor. Please refer to \texttt{src/model/loss} for code implementation.

%% file: 07_Experiments.tex
\section{Sound Matching Evaluation}

Evaluating \modelname{} for sound matching has been a challenging journey, encountering non-convergence in many experiments. Despite this, we successfully established several findings and techniques, presenting a comprehensive synthesizer and platform as a progressive step in the research domain. Our work spanned various synthesizer chains, loss configurations, and neural architectures. While the complexities led to a focus on key findings rather than exhaustive experimental details, we believe this approach illuminates the most promising aspects of our research, fostering further exploration in this field.

Generally, we followed training procedures, hyperparameters and configurations similar to those outlined in~\cite{masuda2021synthesizer}: (1)`P-loss': Training with parameters only; (2)`Synth': Training with parameters loss and gradually shifting to spectral (i.e signal-chain or some other configuration) loss over in-domain data; (3)`Real': Same as (2), but continuing training with spectral loss over out-of-domain data (Nsynth~\cite{Engel2017}). 

\begin{figure}
\setlength{\fboxsep}{0pt}
 \centerline{
 \includegraphics[width=0.55\linewidth]{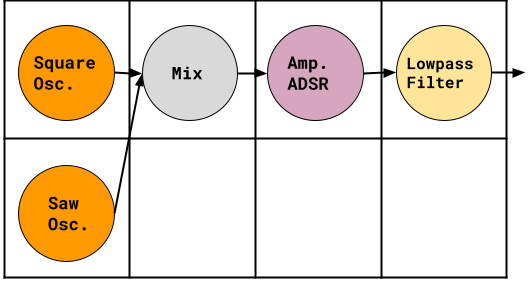}}
 \caption{The chain used for the experiment in Fig. \ref{fig:experiment}, with a sawtooth oscillator, square oscillator, Amplitude ADSR and a Lowpass Filter.}
 \label{fig:reduced_chain}
\end{figure}

\begin{figure}[t]
\centering
\begin{adjustbox}{width=\columnwidth, height=1.9cm, center}
\includegraphics[valign=c,height=1.9cm]{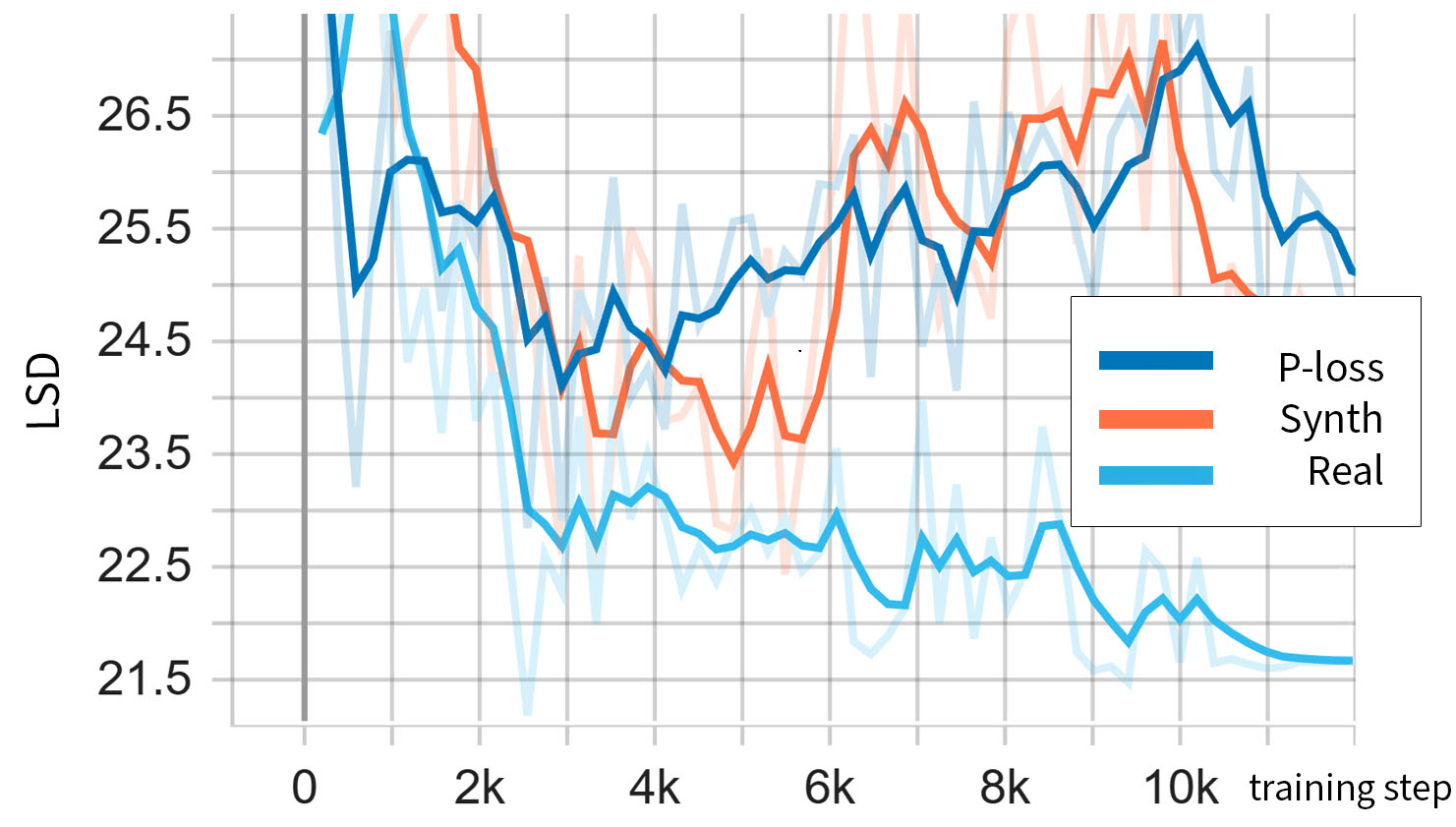}
\end{adjustbox}\vspace{-3mm}
\caption{Training procedures evaluation on NSynth. 'Synth' uses parameter loss until step 2k, then gradually introduces spectral loss (trained on in-domain data) until step 6k. 'Real' mirrors 'Synth' but transitions to out-of-domain data at step 6k. Log-Spectral Distance: \( LSD = \left\lVert (log(S(x)) - log(S(\hat{x}))), \right\rVert_{F} \), \( F \) is the Frobenius norm. The 'Real' procedure shows the efficacy of using out-of-domain data for training, enhancing real-world sound reproductions.}
\label{fig:experiment}
\end{figure}

Using the signal-chain loss solely failed systematically, However, when applied to a relatively basic synthesizer chain (Fig. \ref{fig:reduced_chain}), training with the spectral signal-chain loss after using the parameter loss hinted superior performance over the sole usage of parameters loss on out-of-domain data, which is on par with previous studies~\cite{masuda2021synthesizer, masuda2023synthesizer}. Figure~\ref{fig:experiment} presents an evaluation example using a chain which forms a synth similar to the one presented in~\cite{masuda2021synthesizer}, with a mix of a square and sawtooth oscillators followed by an ADSR amplitude shaper and a filter. 

\begin{figure}[t]
  \centering
  \subfigure[Loss vs. Oscillator Frequency]{\includegraphics[width=0.5\linewidth, height=0.35\linewidth]{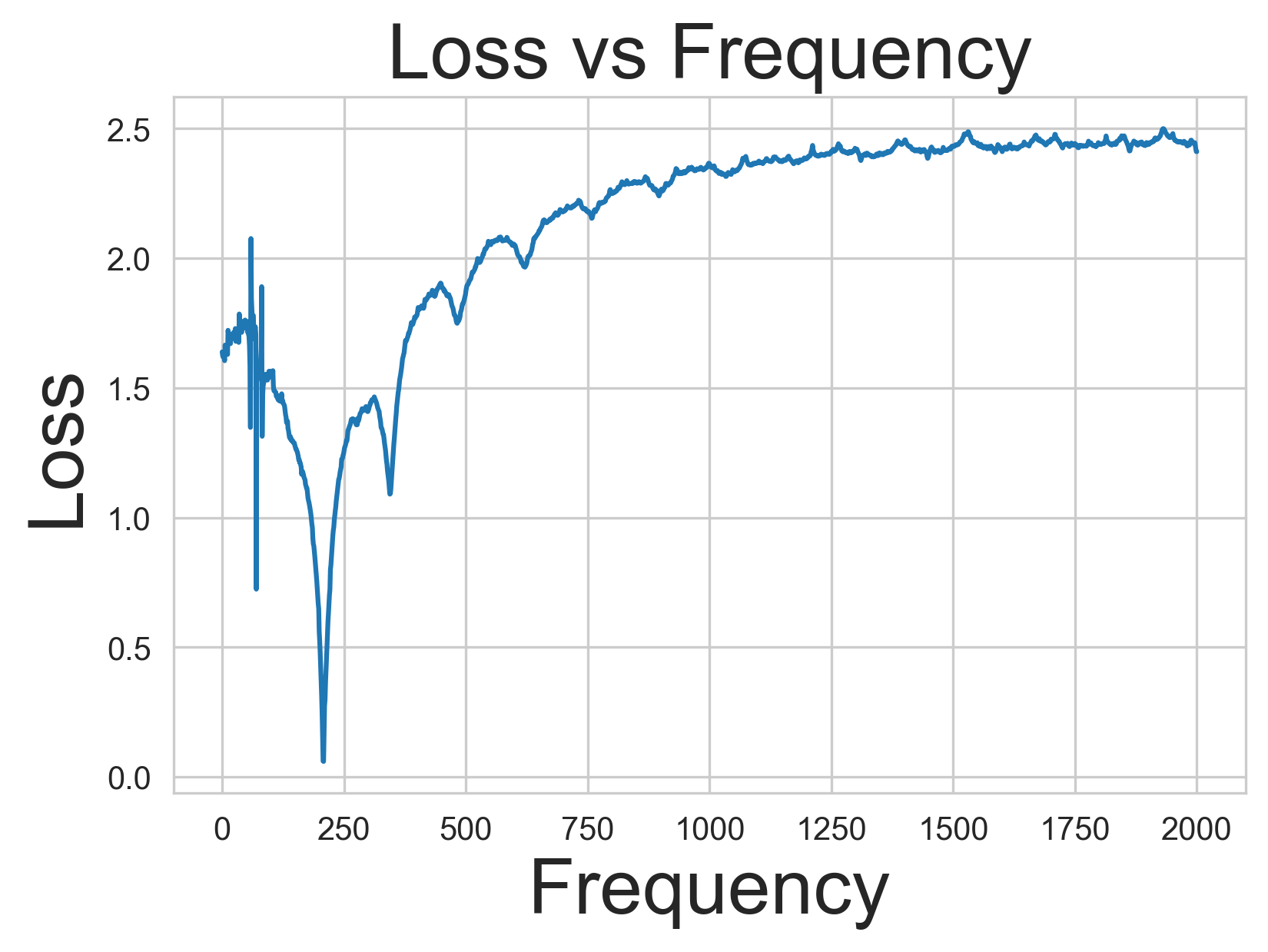}}\hfill
  \subfigure[Loss vs. Modulation Index]{\includegraphics[width=0.5\linewidth, height=0.35\linewidth]{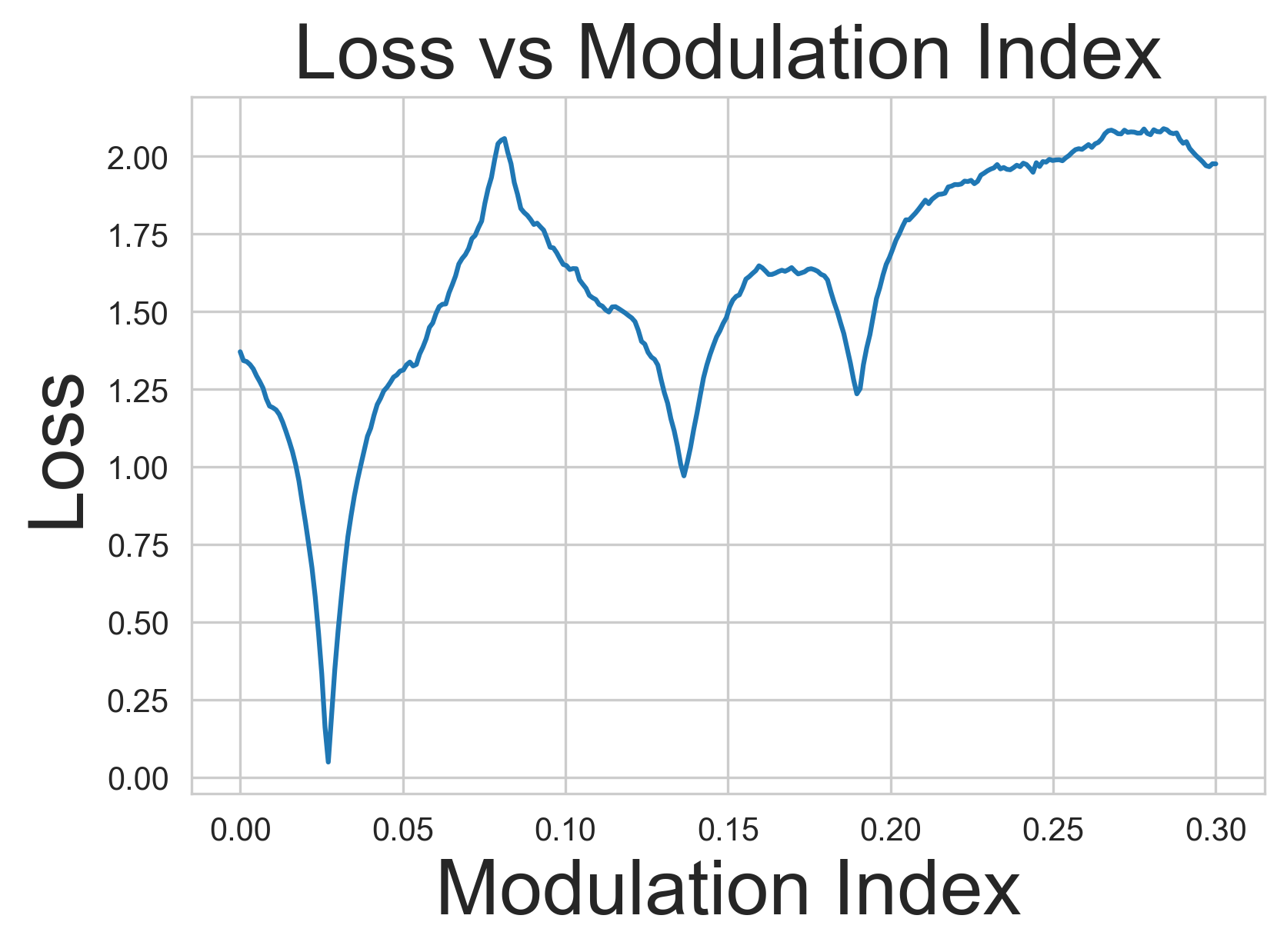}}\\
  \caption{Loss surface for frequency and modulation index, illustrating highly non-convex behavior with many local minima, which pose challenges for optimization. Other synth parameters are fixed to their ground truth.}
  \label{fig:surface_plots}
\end{figure}

We also report that more complex chains utilizing FM modulations refused to converge, adding up to the already hard task of frequency estimation using spectral loss~\cite{turian2020sorry}. The back-propagated gradient from the spectral distance (Eq. \ref{eq:loss_ifs}) for frequency components and the FM modulation index is very abrupt. This phenomenon is conveyed in Fig.~\ref{fig:surface_plots}.


In another experiment, we examined the effectiveness of various spectral processing functions for frequency estimation, adopting the same experimental setup as described in \cite{turian2020sorry}. A deliberately perturbed signal, more distant from a target than a predicted signal, is expected to have a gradient oriented towards the target. This assumption is compared for different loss configurations. As Table\ref{tab:comparison} shows, using the Wasserstein distance on the time axis significantly enhanced the accuracy of frequency estimation for both square and sawtooth waveforms. This outcome defied our initial hypothesis of seeing more benefits when applying Wasserstein distance on the frequency axis. Notably, this approach varies from earlier studies that relied on spectrograms and their log representations for spectral loss 

\textbf{To conclude}, our study indicates that differentiable synthesizers equipped with spectral loss optimization may indeed facilitate sound matching. Yet, achieving high precision in imitating typical sounds remains a formidable challenge. Notably, employing the Wasserstein distance could potentially mitigate the gradient issues encountered in frequency estimation using spectral loss. We anticipate that our platform will catalyze further research in this captivating domain. Moving forward, we propose the exploration of refined audio loss functions, optimization strategies, and alternative neural network architectures to surmount this hurdle.

\begin{table}[t]
\centering
\footnotesize 
\begin{tabular}{|l|c|c|c|}
\hline
\textbf{Loss conf. \textbackslash{} Perturbed dist.} & \textbf{f ± $\varepsilon$} & \textbf{f ± 300 cents} & \textbf{f ± 600 cents} \\ \hline
\multicolumn{4}{|c|}{\textbf{Square waves}} \\ \hline
Spectrogram, $I$             & 0.5   & 0.676 & 0.7   \\ \hline
Mel, $I$     & 0.501 & 0.498 & 0.475 \\ \hline
Spectrogram, Wasserstein time            & 0.48  & \textbf{0.733} & \textbf{0.748} \\ \hline
Spectrogram, Wasserstein freq.            & 0.528 & 0.605 & 0.623 \\ \hline
Mel, Wasserstein time    & 0.488 & 0.466 & 0.452 \\ \hline
Mel, Wasserstein frequency    & 0.494 & 0.696 & 0.645 \\ \hline
\multicolumn{4}{|c|}{\textbf{Sawtooth wave}} \\ \hline
Spectrogram, $I$            & 0.47  & 0.642 & 0.701 \\ \hline
Mel, $I$     & 0.52  & 0.506 & 0.463 \\ \hline
Spectrogram, Wasserstein time           & 0.48  & \textbf{0.712} & \textbf{0.715} \\ \hline
Spectrogram, Wasserstein freq.           & 0.53  & 0.571 & 0.543 \\ \hline
Mel, Wasserstein time   & 0.469 & 0.487 & 0.467 \\ \hline
Mel, Wasserstein frequency   & 0.505 & 0.624 & 0.619 \\ \hline
\end{tabular}
\caption{Comparison of different Loss configurations (modified $S$, $F_\mathrm{l}$, in Eq. \ref{eq:loss_ifs}) and waveshapes across various distances of the pertrubed signal. In a well behaved distance, the prediction
should be closer to the target than the perturbation is. This condition gives us a 0/1 error. Repeated
and averaged over 1000 trials, high accuracies indicate that gradients usually point in the right
direction. f is the predicted wave frequency, $\varepsilon$ represents the local gradient, $I$ is the identity function. 
}
\label{tab:comparison}
\end{table}